\documentclass[english]{npw}
\usepackage[T1]{fontenc}
\usepackage[latin9]{inputenc}
\usepackage{color}
\usepackage{amsmath}
\usepackage{amssymb}
\usepackage{graphicx}
\usepackage{esint}

\makeatletter

\providecommand{\tabularnewline}{\\}

\renewcommand{\vec}{\mathbf}
\newcommand{\pr}[1]{{\sc{\lowercase{#1}}}}

\makeatother

\usepackage{babel}
\begin{document}

\title{Alpha-particle formation and decay rates from Skyrme-HFB wave functions}

\author{D. E. Ward\email{daniel.ward@matfys.lth.se}, B. G. Carlsson, and
S. Åberg\\
\emph{Mathematical Physics, LTH, Lund University, PO Box 118, S-22100
Lund, Sweden}}
\maketitle

\date{\date{}}
\begin{abstract}
$\alpha$ decay is treated microscopically, where the unstable mother
nucleus and residual daughter nucleus are described using HFB wave
functions, obtained with the Skyrme effective interaction. From these
wave functions the amplitude for forming $\alpha$ particles in the
mother nucleus is computed. Two different Skyrme parametrizations
with different pairing properties are compared, and we find good agreement
with experiment for relative decay rates in both cases. The absolute
values of the decay rates are under-estimated. 
\end{abstract}

\section{Introduction}

Nuclear $\alpha$ decay is the process where an unstable nucleus splits
into $^{4}$He ($\alpha$ particle) and a daughter nucleus with two
less protons and two less neutrons. This process is much closer related
to spontaneous fission, i.e., a disintegration of the nucleus into two
nuclei with mass numbers $A\gtrsim60$, than $\beta$ and $\gamma$ decay,
which involve the creation of particles due to the interaction of
the nucleus with weak and electromagnetic fields, respectively.

For superheavy elements (SHE) fission and $\alpha$ decay are the
dominant decay modes. The detection of emitted $\alpha$ particles
has been the principal method of identifying SHEs, created in heavy
ion fusion reactions \cite{Oganessian2007}. By detecting $\alpha$
particles from decays leaving the daughter nucleus in an excited state
in coincidence with the subsequent electromagnetic decay, a first
exploration of nuclear level structure as well as possible identification
by X-ray emission in the $\mathrm{Z=115}$ region was recently made
in the experiment of Ref. \cite{RudolphEtAl2013}. Such $\alpha$ decays
populating excited daughter states occur due to different structure
of mother and daughter nuclei, that hinders the decay to the ground
state \cite{PoggenburgMangRasmussen1969}.

Both fission and $\alpha$-decay modes proceed through classically
forbidden regions of configuration space, i.e., if the motion of the
nucleons is described by some collective coordinates, the nuclear
wave function must tunnel through a potential barrier parametrized
by these coordinates. For fission the collective coordinates usually
characterize the evolution of the overall nuclear shape, while for
$\alpha$ decay one can choose the distance between the center of
mass of the daughter nucleus and the $\alpha$ particle.

The tunneling leads to the huge variation in $\alpha$-decay lifetimes
across the nuclear chart, as the half-life depends roughly exponentially
on the difference between the height of the barrier and the $Q_{\alpha}$ value.
Due to the barrier penetration being the dominant factor determining
the decay probability, this dependence on $Q_{\alpha}$ can be described
using phenomenological $\alpha$-daughter potentials or semi-empirical
formulas. Within such approaches one has been able to reproduce the
overall trends in $\alpha$-decay half-lives using a small number
of free parameters \cite{Qi2009,Zdeb2013}. 

To connect $\alpha$ decay with other nuclear observables, and describe
the hindrance of certain $\alpha$-decay channels, a microscopic model
is needed. In a microscopic description one considers interacting
nucleons, and the formation of an $\alpha$-daughter configuration
from the initial state of the mother nucleus. A predictive description
of both nuclear ground state and $\alpha$-decay properties based
on the same effective interaction is a challenging task for theory.
Most microscopic calculations have been based on either a phenomenological
Woods-Saxon or Nilsson mean field combined with BCS pairing \cite{PoggenburgMangRasmussen1969,Delion1996,Insolia1988},
or a shell-model type description where four nucleons interact with
the daughter nucleus and each other through effective model space
interactions \cite{TonozukaArima1979,Delion2000,Betan2012}. 

In this work we use well-tested effective Skyrme interactions to describe
the nuclear structure. The wave functions of the mother and daughter
nuclei are obtained self consistently using the Hartree-Fock-Bogoliubov
(HFB) method. The Skyrme-HFB approach provides a reasonable description
of ground state observables such as binding energies and densities
throughout the nuclear chart \cite{BenderHeenenReinhard2003}. 

In \cite{Ward2013} we preformed extensive calculations for even even
near-spherical nuclei using the SLy4 effective Skyrme interaction.
In this work we explore if comparable results are obtained using a
different modern Skyrme parametrization, UNEDF1 \cite{Kortelainen2012}.
The connection between the two-particle transfer density, the pairing density and the probability of forming $\alpha$ particles is discussed. Also new for this work is the investigation of the accuracy of the $\delta$-function 
approximation \cite{Rasmussen1963}, which drastically reduces the time to compute the $\alpha$-particle formation amplitude, by comparing with the full calculation in the small $\alpha$-particle limit.   

The contribution is organized as follows. In Sec. \ref{sec:Theory}
the $\alpha$-decay formalism and the nuclear structure model are
briefly introduced. In Sec. \ref{sec:Numerical-application} the convergence
of the numerical results, and the $\delta$-function approximation
are investigated. We discuss the role of pairing correlations in section
\ref{sec:Pairing-correlations}. Results for Po, Ra, and Rn isotopes
are compared with experiment in Sec. \ref{sec:Results}. Finally,
in Sec. \ref{sec:Conclusions} we present our conclusions.

\section{Theory\label{sec:Theory}}

We use the same decay formalism and Skyrme-HFB nuclear structure model
as in the previous work \cite{Ward2013}, where several aspects of
the approach and the introduced approximations are discussed in detail.
Here we recapitulate only the ingredients necessary for the discussion
of the current results.

\subsection{Decay formalism}

$\alpha$-decay is described in the microscopic $R$-matrix approach
\cite{Lovas1998,Zeh1963,Mang1964}. The amplitude of the decay process
depends on the overlap of the mother $(M)$ nucleus with a daughter
$(D)$ and $\alpha$ cluster, called the\emph{ formation amplitude}
\cite{Lovas1998},

\textcolor{black}{
\begin{equation}
\begin{split} & g_{L}(r_{\alpha D})=\int\mathcal{A}\left[\Phi_{J}^{\left(D\right)}(\xi_{D}),\Phi_{0}^{\left(\alpha\right)}(\xi_{\alpha})Y_{L}\left(\hat{r}_{\alpha D}\right)\right]_{IM}^{*}\\
 & \times\Psi_{IM}^{(M)}(\xi_{M})d\xi_{D}d\xi_{\alpha}d\hat{r}_{\alpha D},
\end{split}
\label{eq:FormationAmplitudeIntegral-1-2}
\end{equation}
where $L$ is the angular momentum, }$\Psi_{IM}^{(M)}(\xi_{M})$,\textcolor{black}{$\Phi_{J}^{\left(D\right)}(\xi_{D})$,}
and $\Phi_{0}^{\left(\alpha\right)}(\xi_{\alpha})$ are the mother,
daughter and $\alpha$ particle wave functions, respectively, $\xi_{i}$
are intrinsic coordinates for nucleus $i$, and $\mathbf{r}_{\alpha D}$
is the distance between the centers of mass of the daughter nucleus
and $\alpha$ particle. In this work we consider the g.s. to g.s.
$\alpha$ decay in even-even nuclei implying $L=J=I=0$.

The decay width is given by \cite{Lovas1998},

\begin{equation}
\Gamma(r_{c})=2\gamma_{0}^{2}(r_{c})P_{0}(Q_{\alpha},r_{c}),\label{eq:RmatrixGamma}
\end{equation}
where $P_{0}$ is the Coulomb penetrability, and $\gamma_{0}$ is
the reduced width, \textcolor{black}{
\begin{equation}
\gamma_{0}^{2}(r_{c})=\frac{\hbar^{2}}{2\mu r_{c}}r_{c}^{2}g_{0}^{2}(r_{c}),\label{eq:ReducedWidth}
\end{equation}
}with the reduced mass $\mu$. The penetrability depends strongly
on $Q_{\alpha}$. A variation of $\pm1~\mathrm{MeV}$ can produce a variation of several orders of magnitude in the penetrability. To avoid this we use the experimental value, $Q_{\alpha}^{\mathrm{exp}}$. 

The reduced width and the penetrability entering Eq. (\ref{eq:RmatrixGamma})
both depend on the matching radius $r_{c}$, that should be chosen
beyond the range of inter-cluster nuclear forces. If the formation
amplitude has a tail proportional to an outgoing Coulomb wave function
corresponding to the correct energy $Q_{\alpha}^{\mathrm{exp}}$, the decay
width $\Gamma$ becomes constant for large $r_{c}$. For an approximate
nuclear structure model there might thus be some $r_{c}$ dependence
of $\Gamma$.

\subsection{Nuclear structure model}

The g.s. wave functions of mother and daughter nuclei are obtained
from spherically symmetric solutions to the HFB equations, solved
in a large spherical oscillator basis using an extended version of
the program \pr{HOSPHE}(v1.02) \cite{Carlsson2010p2}. In the particle-hole
(p-h) channel the effective Skyrme forces SLy4 and UNEDF1 are used. 

In the particle-particle (p-p) channel an effective density-dependent
contact interaction \cite{Dobaczewski2004} is used,

\begin{equation}
V_{pair}^{q}(\mathbf{r},\mathbf{{r}')}=V_{q}\left[1-\beta\frac{{\rho(\mathbf{\mathbf{{r)}}}}}{\rho_{c}}\right]\delta(\mathbf{{r}-\mathbf{{r}'}}),\; q=p,n,\label{eq:pairingPot}
\end{equation}
where the parameters $V_{q}$ and $\beta$ determine the strength
and density dependence of the interaction, and $\rho_{c}$ is the
saturation density of nuclear matter, taken as $\rho_{c}=0.16\:\mathrm{fm}^{-3}$.
The interaction is regularized by a truncation in the equivalent spectra
\cite{Stoitsov2005} at 60 MeV. To avoid a collapse of the pairing
in nuclei close to closed shells an approximate version of the Lipkin-Nogami
method \cite{Stoitsov2005} is used. The Skyrme force SLy4 is used
in combination with volume-type ($\beta=0$), and surface-type pairing
($\beta=1$), with the pairing strengths used in \cite{Ward2013}.
For UNEDF1 we use the mixed pairing ($\beta=\frac{1}{2}$) obtained
in the simultaneous fit of the p-h and p-p effective interactions
performed in \cite{Kortelainen2012}. 

We use the standard approximation for the $\alpha$ particle wave
function \cite{Lovas1998,Delion2010}, 

\textcolor{black}{
\begin{equation}
\begin{split} & \Phi_{00}^{\left(\alpha\right)}(\vec{r}_{\pi},\vec{r}{}_{\nu},\vec{r}{}_{\alpha},s_{1},s_{2},s_{3},s_{4})\\
 & =\left(\frac{4}{b_{\alpha}^{3}\sqrt{\pi}}\right)^{3/2}e^{-\frac{r_{\pi}^{2}+r_{\nu}^{2}+r_{\alpha}^{2}}{2b_{\alpha}^{2}}}\times\left(\frac{1}{\sqrt{4\pi}}\right)^{3}\\
 & \times[\chi_{\frac{1}{2}}(s_{1}),\chi_{\frac{1}{2}}(s_{2})]_{00}[\chi_{\frac{1}{2}}(s_{3}),\chi_{\frac{1}{2}}(s_{4})]_{00},
\end{split}
\label{eq:alphaWF}
\end{equation}
}where $b_{\alpha}$ is a size parameter, chosen as $b_{\alpha}=1.41$
fm, $s_{1},s_{2}$ are spin coordinates for the two protons, and $s_{3},s_{4}$
for the two neutrons. The coordinates $\vec{r}_{\pi},\vec{r}_{\nu},\vec{r}{}_{\alpha}$
are related to the coordinates of the two protons $\vec{r}{}_{1},\vec{r}{}_{2}$
and of the two neutrons $\vec{r}{}_{3},\vec{r}{}_{4}$ through, $\vec{r}_{\pi}=(\vec{r}_{1}-\vec{r}_{2})/\sqrt{2}$,
$\vec{r}_{\nu}=(\vec{r}_{3}-\vec{r}_{4})/\sqrt{2}$, $\vec{r}_{\alpha}=\frac{1}{2}\left(\vec{r}_{1}+\vec{r}_{2}-\vec{r}_{3}-\vec{r}_{4}\right)$.
The center of mass coordinate for the four nucleons is, $\vec{R}=\frac{1}{4}\left(\vec{r}_{1}+\vec{r}_{2}+\vec{r}_{3}+\vec{r}_{4}\right).$

\subsection{Calculation of the formation amplitude}

The formation amplitude (\ref{eq:FormationAmplitudeIntegral-1-2})
is obtained by expanding the mother nucleus in terms of the daughter
and valence nucleons \cite{Lovas1998}, and integrating over daughter
coordinates,

\textcolor{black}{
\begin{equation}
\begin{split} & g_{0}(R)=\frac{1}{\sqrt{4\pi}}\int\Phi_{00}^{\left(\alpha\right)*}(r_{\pi},r_{\nu},r_{\alpha})\Phi^{(v)}(\vec{r}_{1},\vec{r}_{2},\vec{r}_{3},\vec{r}_{4})d\xi_{\alpha}d\hat{R},\end{split}
\label{eq:FormationAmplitudeIntegral-1-1}
\end{equation}
}where $\mathbf{R}$ is the center of mass coordinate for the nucleons
of the $\alpha$ particle, and $\Phi^{(v)}=\sqrt{8}\Phi^{\left(v_{p}\right)}\left(\vec{r}_{1},\vec{r}_{2}\right)\Phi^{\left(v_{n}\right)}\left(\vec{r}_{3},\vec{r}_{4}\right)$
is the valence nucleon wave function. The factor of $\sqrt{8}$ is
the renormalization of a wave function obtained for the coordinates
$\vec{r}_{1},\vec{r}_{2},\vec{r}_{3},\vec{r}_{4}$, when used as a
wave function in the coordinates $\vec{r}_{\pi},\vec{r}_{\nu},\vec{r}{}_{\alpha},\vec{R}$
\cite{Eichler1965}. A few approximations are made to get this simpler
expression, discussed in \cite{Ward2013}. If the daughter nucleus
is heavy compared to the $\alpha$ particle, and their relative distance,
$R$, is large so that they have little spatial overlap, the errors
introduced by these approximations are estimated to be small. 

The two-proton, $q=\pi$, and two-neutron parts, $q=\nu$, of the
valence nucleon wave function are given by,

\begin{equation}
\begin{split}\Phi^{\left(v_{q}\right)}\left(\vec{r}_{a},\vec{r}_{b}\right) & =\frac{1}{2}\sum_{lj}\sum_{nn'}\hat{j}B_{lj;nn'}^{q}\\
 & \times\mathcal{A}\{\left[\phi_{nlj}\left(\vec{r}_{a}\right),\phi_{n'lj}\left(\vec{r}_{b}\right)\right]_{00}\},
\end{split}
\end{equation}
where $\phi_{nlj}$ are harmonic oscillator eigenfunctions with $n$
nodes, orbital and total angular momentum $l$ and $j$, respectively.
Since proton-neutron pairing is not considered the HFB wave functions
splits into proton and neutron parts. The expansion coefficients
$B^{q}$ are given by the overlap of the corresponding proton or
neutron parts of the daughter HFB vacuum with two particle-annihilation
operators acting on the mother HFB vacuum, 
\begin{equation}
B_{k,k'}^{q}=\langle D^{q}|a_{k'}a_{k}|M^{q}\rangle.\label{eq:symXkk-1-1-1-1}
\end{equation}

The formation amplitude (\ref{eq:FormationAmplitudeIntegral-1-1})
is subsequently evaluated by transforming the valence wave function
to relative and center of mass coordinates using Talmi-Moshinsky oscillator
brackets \cite{Kamuntavicius2001}.

\section{Numerical application\label{sec:Numerical-application}}

\subsection{Convergence\label{sub:Convergence}}

\begin{figure}
\includegraphics[width=1\columnwidth]{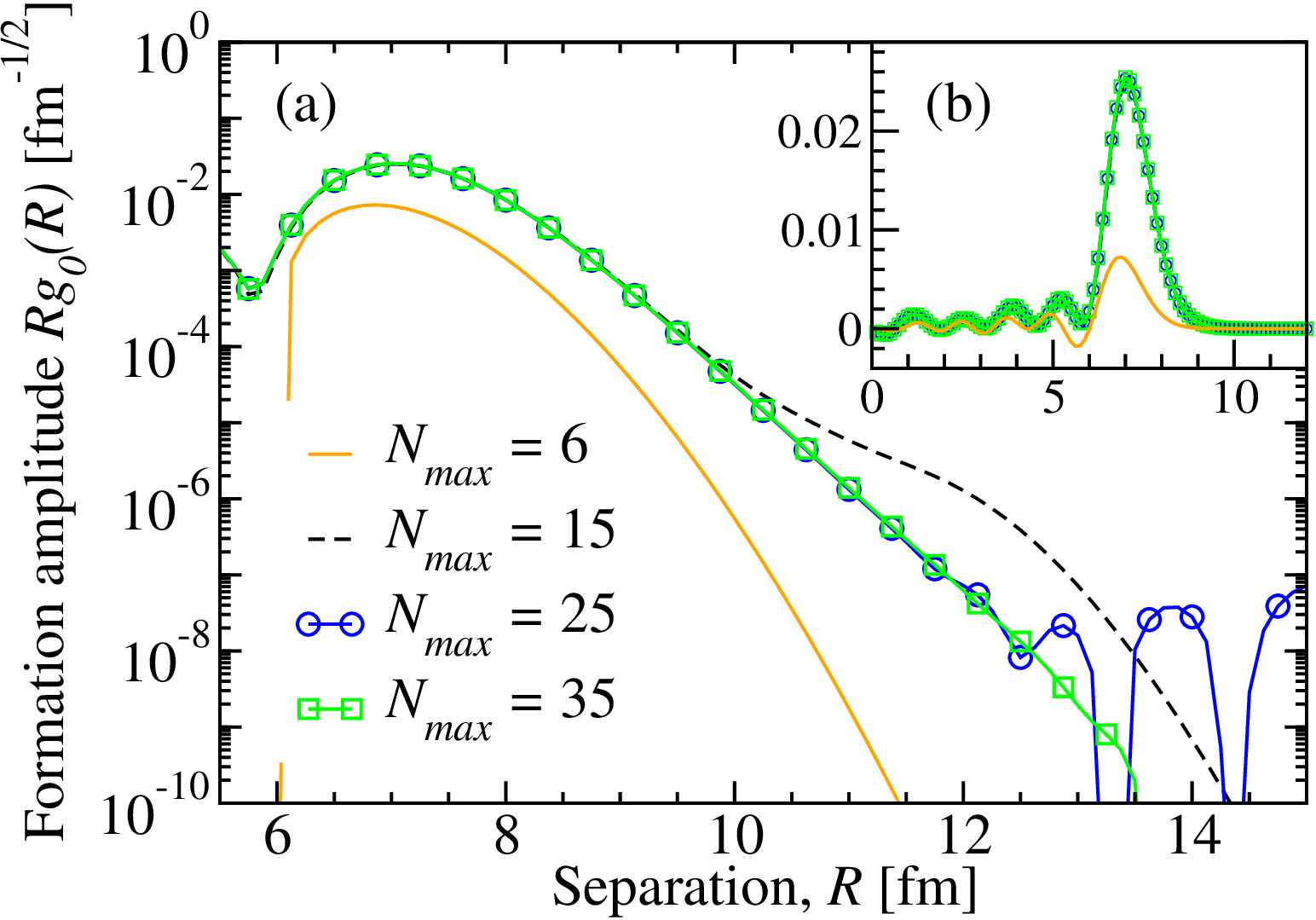}\caption{Convergence of the formation amplitude. (a) shows the tail of the
formation amplitude in logarithmic scale. (b) shows the formation
amplitude in linear scale, only the $N_{max}=6$ results differ from
the rest in this scale. \label{fig:ConvFA}}
\end{figure}

The matching radius, $r_{c}$, in the $R$-matrix decay width expression
(\ref{eq:RmatrixGamma}) must be chosen in the region where nuclear
forces between the alpha particle and the daughter nucleus are negligible.
In the case of $^{212}$Po this means \cite{Ward2013} that the formation
amplitude should be converged for separations $R\leq10$ fm. As seen
in Fig. \ref{fig:ConvFA}
 this is fulfilled when $N_{max}\geq20$, where the oscillator basis
consists of all major shells up to and including $N_{max}$. In this
work we use, unless otherwise noted, $N_{max}=30$ throughout.

\subsection{$\alpha$ particle wave function - small size limit}

The evaluation of the formation amplitude (\ref{eq:FormationAmplitudeIntegral-1-1}),
involves the computationally expensive evaluation of overlap integrals
in relative coordinates. Within the $\delta$-function approximation
\cite{Rasmussen1963}, these computations are avoided, reducing the
computational time by at least two orders of magnitude. The error
introduced vary from nuclei to nuclei, partly due to that the contribution
from high-$j$ orbitals is exaggerated \cite{Rasmussen1963}, and
the approximation has generally not been used in recent works. To
better understand how the errors are introduced, we investigate the
formation amplitude in the small $\alpha$-particle limit.

The $\delta$-approximation consists in assuming the valence wave
function in the coordinates $r_{\pi},r_{\nu},r_{\alpha}$ be constant
over the size of the $\alpha$ particle,

\begin{align}
\Phi^{(v)}(\vec{r}_{\pi},\vec{r}{}_{\nu},\vec{r}{}_{\alpha},\vec{R}) & \approx\Phi^{(v)}(0,0,0,\vec{R})\nonumber \\
 & =\sqrt{8}\Phi^{\left(v_{\pi}\right)}\left(\vec{R},\vec{R}\right)\Phi^{\left(v_{\nu}\right)}\left(\vec{R},\vec{R}\right).
\end{align}
Inserting into Eq. (\ref{eq:FormationAmplitudeIntegral-1-1}) gives,

\begin{equation}
\begin{split} & g_{0}(R)\approx\int\Phi_{00}^{\left(\alpha\right)*}(\vec{r}_{\pi},\vec{r}{}_{\nu},\vec{r}{}_{\alpha})d\xi_{\alpha}\\
 & \times\sum_{s_{1}\dots s_{4}}\int\Phi_{00}^{\left(\alpha\right)*}(s_{1},\dots,s_{4})Y_{00}^{*}(\hat{R})\\
 & \times\Phi^{(v)}(0,0,0,\vec{R};s_{1},\dots,s_{4})d\hat{R},
\end{split}
\label{eq:formampApprox1}
\end{equation}
where the $\alpha$ particle wave function is separated into spin
and position coordinates, and the summation over spins is shown explicitly.
The first integral gives the prefactor, 

\begin{equation}
A(b_{\alpha})\equiv\int\Phi_{00}^{\left(\alpha\right)*}(\vec{r}_{\pi},\vec{r}{}_{\nu},\vec{r}{}_{\alpha})d\xi_{\alpha}=\left(2b_{\alpha}\sqrt{\pi}\right)^{9/2}.
\end{equation}
This factor differs slightly from the one in \cite{Rasmussen1963}.

Performing the summation over spins, and the angular integration,
the formation amplitude becomes,
\begin{equation}
\begin{split} & g_{0}(R)\approx\sqrt{32\pi}A(b_{\alpha})\langle D^{\pi}|M^{\pi}\rangle\langle D^{\nu}|M^{\nu}\rangle\\
 & \times\kappa_{\pi}^{DM}(\mathbf{R})\kappa_{\nu}^{DM}(\mathbf{R}),
\end{split}
\label{eq:formampApprox1-1-1}
\end{equation}
where $\kappa_{q}^{DM}(\mathbf{R})$ is the two-particle transfer
density, here defined as, 
\begin{equation}
\kappa_{q}^{DM}(\mathbf{R})\equiv\frac{\langle D^{q}|a(\mathbf{R},\downarrow)a(\mathbf{R},\uparrow)|M^{q}\rangle}{\langle D^{q}|M^{q}\rangle},\; q=\pi,\nu.\label{eq:kappaDM}
\end{equation}

\begin{figure}
\includegraphics[width=1\columnwidth]{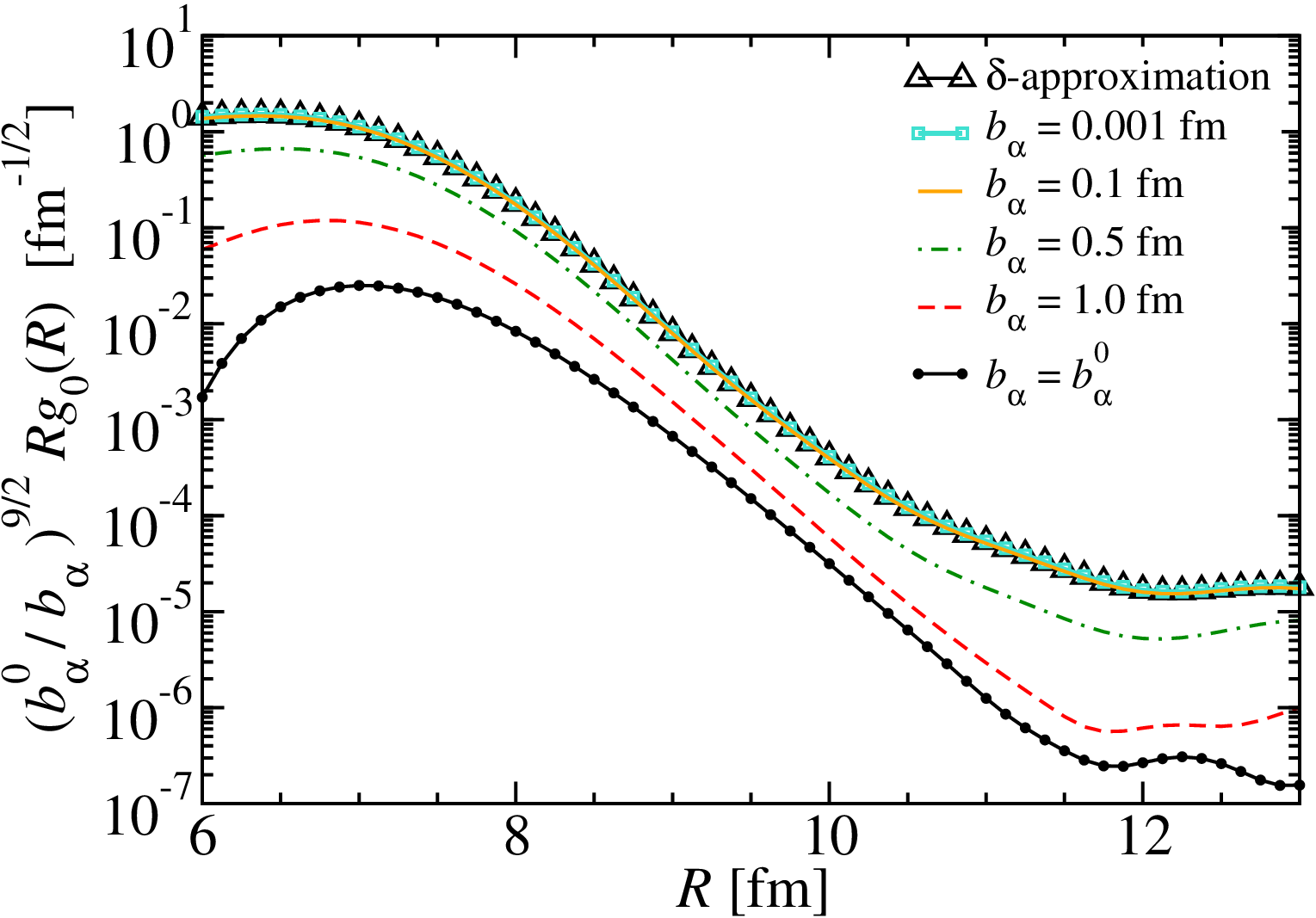}\caption{Formation amplitude for $^{212}$Po calculated for $\alpha$ particle
wave functions of different sizes $b_{\alpha}$, and rescaled, compared
to the result of the $\delta$-function approximation.\label{fig:DeltaApprox} }
\end{figure}

The formation amplitude in the $\delta$-function approximation, using
Eq. (\ref{eq:formampApprox1-1-1}), is compared to the result of the
full calculation, Eq. (\ref{eq:FormationAmplitudeIntegral-1-1}),
for the $\alpha$ decay of $^{212}$Po in Fig.~\ref{fig:DeltaApprox}.
 The HFB wave functions are obtained with SLy4 and volume pairing,
and oscillator shells up to $N_{max}=20$ are included. To investigate
for what size of $\alpha$ particle the assumption of the $\delta$-function
approximation is valid, the results of full calculations using an
$\alpha$ particle wave function with size parameter $b_{\alpha}$,
smaller than the standard value of $b_{\alpha}^{0}=1.41\:\mathrm{fm}$,
are also shown. The results are scaled by the factor $(b_{\alpha}^{0}/b_{\alpha})^{9/2}$,
so that a constant valence-nucleon wave function will produce the
same formation amplitude regardless of $b_{\alpha}$. 

The $\delta$-function approximation over-estimates the formation
amplitude by a factor of 10 or more. From the results of calculations
with decreasing $b_{\alpha}$ we note that this overestimation is
due to the neglect of the decrease of the valence-nucleon wave function
with increasing distance between nucleons $r_{\pi},r_{\nu},r_{\alpha}$.
As the volume in which the valence-nucleon wave function is sampled,
controlled by the parameter $b_{\alpha}$, is decreased the results
tend to the results of the $\delta$-function approximation. 

We conclude that the $\delta$-approximation is not sufficiently accurate.
More work is needed to find a suitable approximation to reduce the
computational time. For the remaining calculations we shall use the
full expression (\ref{eq:FormationAmplitudeIntegral-1-1}) for the
formation amplitude.

\section{Pairing correlations\label{sec:Pairing-correlations}}

The formation amplitude depends strongly on the magnitude of the pairing
correlations. For BCS type calculations a strong dependence on the
paring gap was demonstrated in e.g. \cite{Delion1996}, and a similar
dependence was observed in the HFB case \cite{Ward2013}. 

\begin{figure}
\includegraphics[width=1\columnwidth]{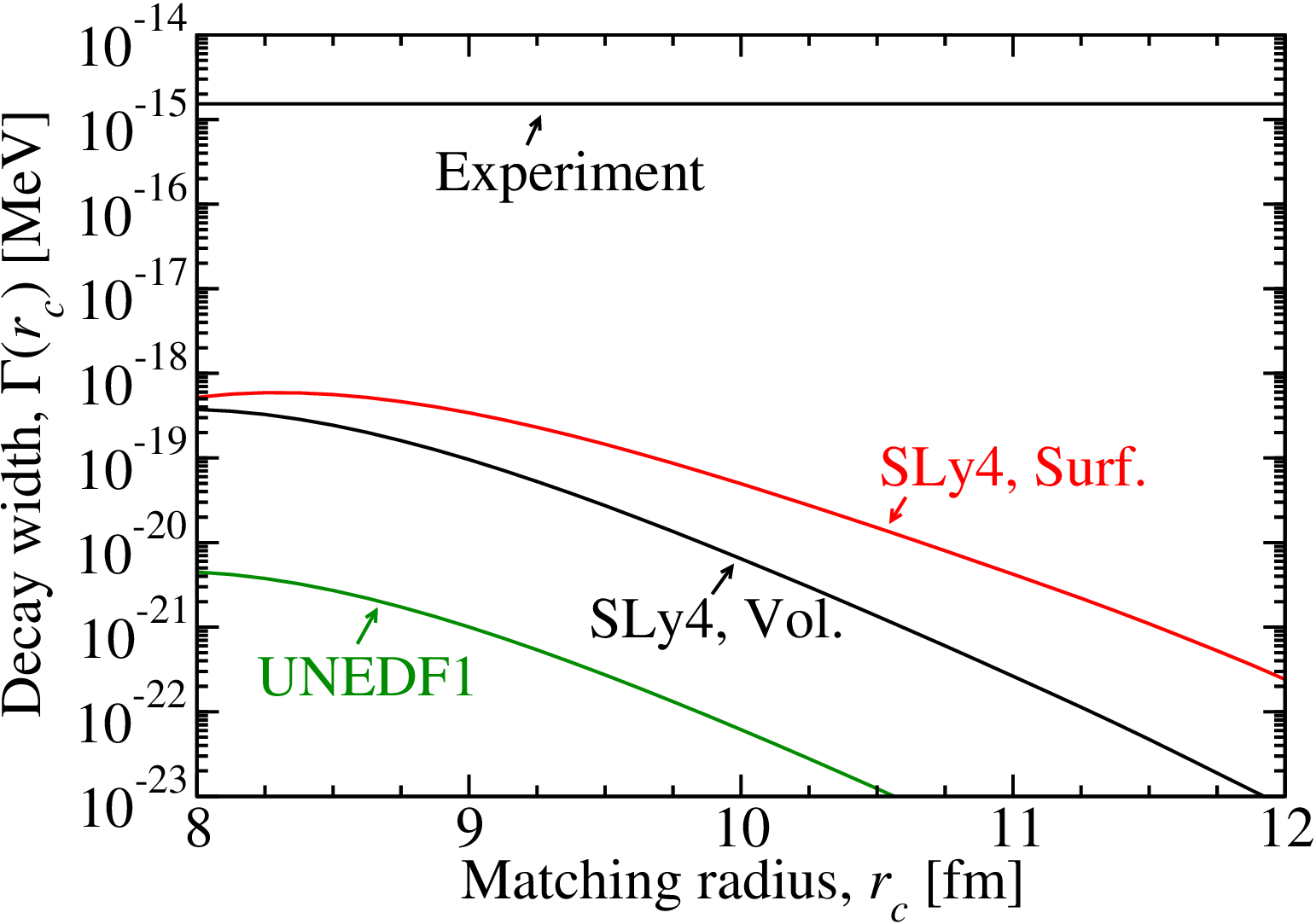}\caption{$R$-matrix decay width for $^{212}$Po, obtained using different
effective interactions. \label{fig:RmatGamma}}
\end{figure}

The $R$-matrix decay width for $^{212}$Po calculated using different
effective Skyrme and pairing interactions is shown in Fig. \ref{fig:RmatGamma}.
 As can be seen in the figure all considered interactions produce
too small decay widths to reproduce experimental data. SLy4 with a
surface pairing force yields the largest values, whereas the UNEDF1
produces the smallest decay width. The slope of $\Gamma(r_{c})$ as
a function of matching radius $r_{c}$ indicates that the model predicts
an $\alpha$ particle more bound to the daughter compared to experiment. 

While both SLy4 results are obtained using a pairing that gives lowest
quasi-particle energies, $E_{q}^{\mathrm{min}}$, close to the three-point
mass differences, $\Delta_{p}^{\mathrm{exp}}(^{212}\mathrm{Po})=0.765$ MeV,
and $\Delta_{n}^{\mathrm{exp}}(^{212}\mathrm{Po})=0.766$ MeV, the UNEDF1 produces
smaller pairing, $E_{p}^{\mathrm{min}}(^{212}\mathrm{Po})=0.144$ MeV, $E_{n}^{\mathrm{min}}(^{212}\mathrm{Po})=0.111$
MeV. This pairing produces the much smaller decay width in this case. 

\begin{figure}
\includegraphics[bb=21bp 25bp 721bp 523bp,clip,width=1\columnwidth]{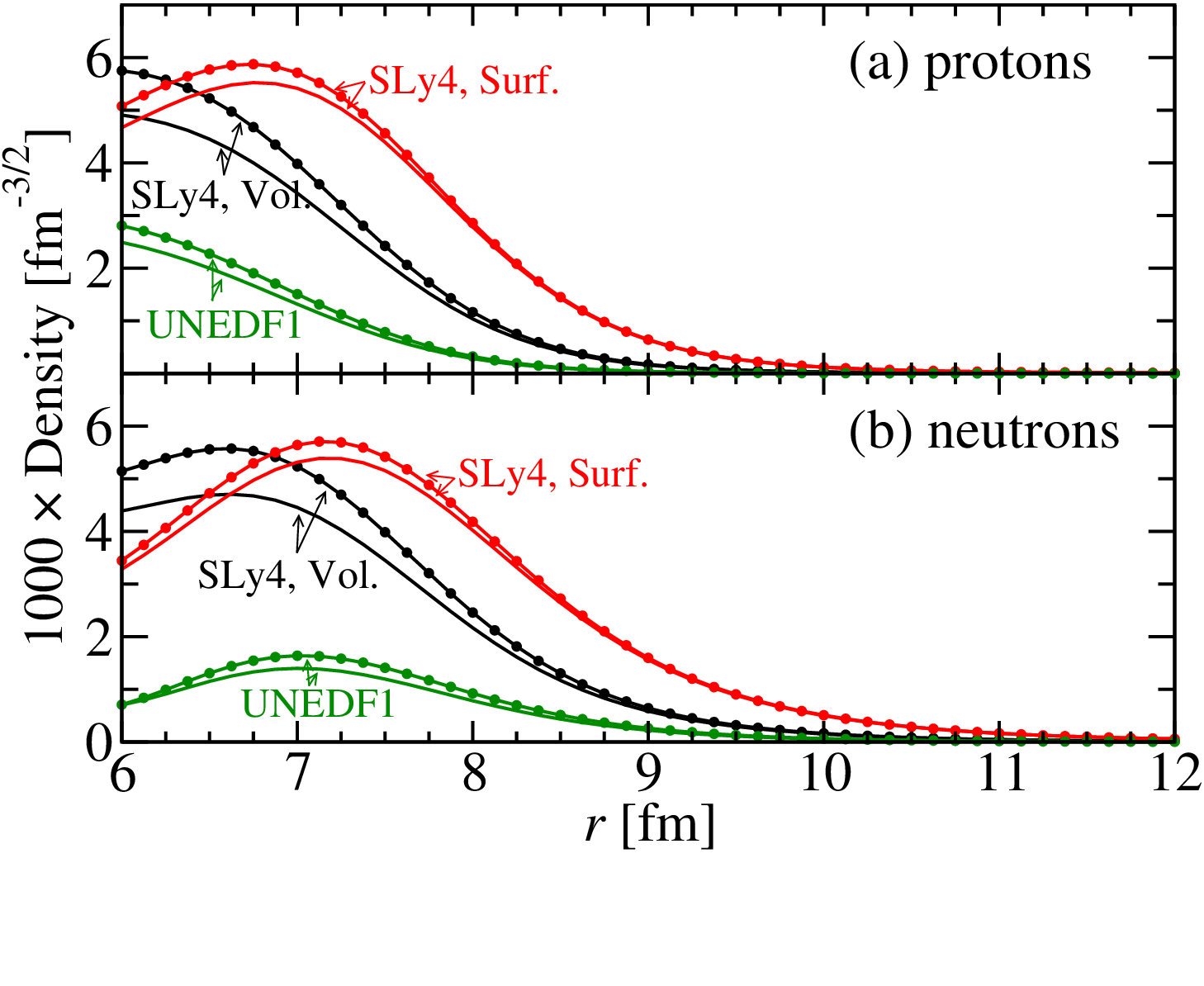}\caption{Pairing density (solid lines), and two-nucleon transfer density (dotted
lines) for $^{212}$Po.\label{fig:kappaDens-1}}
\end{figure}

The connection between the pairing and the formation amplitude can
be seen more clearly in the $\delta$-function approximation, Eq.
(\ref{eq:formampApprox1-1-1}). In this approximation the formation
amplitude is proportional to the product of the pair-transfer densities
$\kappa_{p}^{DM}(\mathbf{R})\kappa_{n}^{DM}(\mathbf{R})$. Fig. \ref{fig:kappaDens-1}
 compares the pairing density, $\kappa_{q}(\mathbf{r})=\langle M^{q}|a(\mathbf{r},\downarrow)a(\mathbf{r},\uparrow)|M^{q}\rangle$,
and the two-nucleon transfer densitiy, $\kappa_{q}^{DM}(\mathbf{r})$,
Eq. (\ref{eq:kappaDM}), calculated in the three model parametrizations
for $^{212}$Po. It is interesting to note that for all three cases
the transfer density seem to be well approximated by the pairing density
for large radii. The surface pairing has the largest tail of the pair
density, while the UNEDF1 pairing density is overall much smaller
due to the smaller pairing. This behavior explains qualitatively the
$R$-matrix decay widths in Fig. \ref{fig:RmatGamma}.

\section{Results for $\alpha$-decay properties of Po, Rn, Ra isotopes\label{sec:Results}}

Calculations for $\alpha$-decay widths of all even-even near-spherical
$\alpha$ emitters included in the compilation of experimental data
of \cite{Akovali1998} were performed using the SLy4 Skyrme force
in \cite{Ward2013}. Here we focus on the results for the 27 near-spherical Po, Rn, and Ra isotopes, and compare with results using
the UNEDF1 effective interaction.

To compare the microscopic $\alpha$-decay properties for several
nuclei we investigate the reduced width, $\gamma^{2}(r_{c})$, Eq. (\ref{eq:ReducedWidth}),
at $r_c = r_t$, with the touching radius $r_{t}=r_0\left[(A-4)^{1/3}+4^{1/3}\right],$
where $A$ is the mass number of the mother nucleus and $r_0 = 1.2~\mathrm{fm}$. $r_{t}$ gives
the approximate separation at the $\alpha$-daughter nucleus touching
configuration. 

For g.s. to g.s. $\alpha$ decay of even spherical nuclei, the experimental
reduced width is defined as,

\begin{equation}
\gamma_{\mathrm{exp}}^{2}(r_{t})=\frac{\Gamma_{\mathrm{exp}}}{2P_{0}(Q_{\alpha}^{\mathrm{exp}},r_{t})},
\end{equation}
where $\Gamma_{\mathrm{exp}}=\hbar\:\mathrm{ln}\:2/T_{1/2}^{\mathrm{exp}}$ is the partial
decay width for the g.s. to g.s. decay channel and $P_{0}$ is the Coulomb
penetrability.

Tab. \ref{tab:MeanStd}
\begin{table}
\caption{Mean, $\mathcal{M}$ , and standard deviation, $\sigma$, of $\mathrm{log_{10}[\gamma^{2}(r_{t})/\gamma_{\mathrm{exp}}^{2}(r_{t})]}$
for the 27 studied Po, Rn and Ra isotopes.\label{tab:MeanStd} }
\centering

\begin{tabular}{ccc}
\hline 
Model & $\mathcal{M}$ & $\sigma$\tabularnewline
\hline 
SLy4, Volume pairing & -3.7897 & 0.3162\tabularnewline
SLy4, Surface pairing & -3.1409 & 0.2213\tabularnewline
UNEDF1 & -5.7080 & 0.2278\tabularnewline
UDL & 0.2111 & 0.3788\tabularnewline
\hline 
\end{tabular}
\end{table}
 shows the mean $\mathcal{M}$ and standard deviation $\sigma$
of the logarithm of theoretical reduced widths divided by the corresponding
experimental value for the 27 isotopes, 
\begin{equation}
\mathrm{log_{10}}\left[\frac{\gamma^{2}(r_{t})}{\gamma_{\mathrm{exp}}^{2}(r_{t})}\right]=\mathrm{log_{10}}\left[\frac{\Gamma(r_{t})}{\Gamma_{\mathrm{exp}}}\right].
\end{equation}
For comparison results from the semi-empiric formula, Universal Decay
Law (UDL), of Ref \cite{Qi2009} are included. One notes that the
microscopic models produce reduced widths which are on average 3 or
more orders of magnitude smaller than experiment, whereas the UDL
produces on the average a good agreement with data. The spread, $\sigma$,
around this mean is, on the other hand, smaller for the microscopic
models, with the smallest deviation for the SLy4 + Surface pairing
and the UNEDF1 effective interactions. 

\begin{figure}
\includegraphics[width=1\columnwidth]{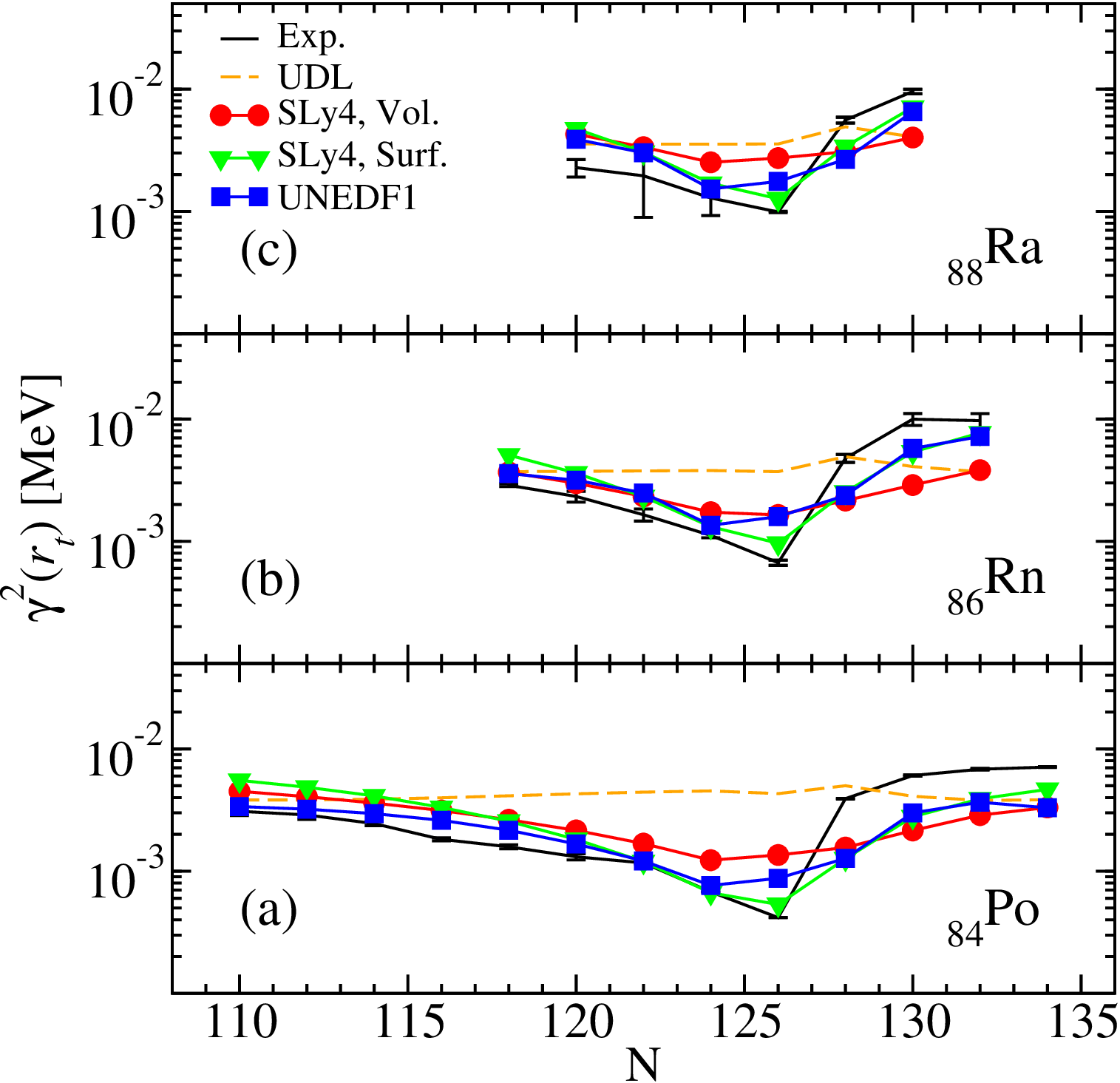}\caption{Reduced widths at the touching radius, as a function of the mother
nucleus neutron number. The error bars show experimental values, the
dashed line results from the UDL formula. The circles, triangles and
squares show results from microscopic HFB calculations using different
effective interactions. The microscopic results are normalized with
the constant factor $10^{-\mathcal{M}}$.\label{fig:Reduced-widths}}
\end{figure}

The reduced widths are compared with experiment in Fig. \ref{fig:Reduced-widths}.
 For each of the microscopic models the results are normalized with
the constant $10^{-\mathcal{M}}$. From the figure it is apparent
that the normalized microscopic results (circles, triangles, and squares)
follow the variation of $\gamma_{\mathrm{exp}}^{2}(r_{t})$ better than the
semi-empiric model (dashed line), that does not take into account
the nuclear structure.

Reduced $\alpha$-decay widths along the Po, Rn and Ra isotope chains
show a similar behavior: As the neutron number is increased, the reduced
width decreases rather smoothly until the neutron shell closure at
N = 126, after which the experimental value increases drastically
across the shell gap. For $\mathrm{N}\geq128$ the reduced widths
again increase in a smooth way as N is increased. None of the models
fully capture the magnitude of the increase in reduced width when
crossing the gap. The best agreement is for SLy4, with surface type
pairing. For all microscopic models most of the variation with neutron
number is reasonably reproduced.

This shows that although the magnitude of the calculated formation
amplitudes are much too small, the missing nuclear structure effects
needed to reproduce data is roughly proportional to the HFB results
for all three considered model parametrizations.

The reduced width $\gamma^{2}(r_{c})$ depends on the matching radius $r_c$, and decreases with increasing values of this parameter, as the tail of the formation amplitude decreases inside the Coulomb barrier. As noted in Sec. \ref{sec:Pairing-correlations} the decrease of the microscopic reduced width is more rapid than the corresponding experimental values, causing the theoretical decay widths $\Gamma(r_c)$, shown in Fig. \ref{fig:RmatGamma}, to decrease with $r_c$. This implies that the HFB results presented in Tab. \ref{tab:MeanStd} and Fig. \ref{fig:Reduced-widths} depend on the choice of $r_c$. To test the sensitivity of the results to $r_c$ we use a scaled touching radius $r_t^\prime$ with $r_0 = 1.5~\mathrm{fm}$. While the mean values decrease by about 2 orders of magnitude, the relative values still show  a good agreement with data, reflected in small changes in the standard deviations $\sigma$.

\section{Conclusions\label{sec:Conclusions}}

The magnitude of the reduced $\alpha$-decay widths turned out to
be on average 5.7 orders of magnitude too small compared to experiment
for the UNEDF1, whereas the SLy4 wave functions produced 3-4 orders
too small amplitudes. The much smaller UNEDF1 values are mainly due
to the weaker pairing.

The three considered model parametrizations have quite different pairing
properties. In spite of this all three provide relative decay rates
that are in good agreement with data. The good agreement for the normalized
widths indicates that the missing effects needed to reproduce absolute
values of experimental decay rates seem to be roughly proportional
to the Skyrme-HFB results.

To describe these missing effects it is likely that one has to go
beyond the HFB level and introduce other types of correlations. Such
a model would require more complicated wave functions, that might
make the calculation of the formation amplitudes intractable. A great
reduction in computer time can be achieved if the computation of overlap
integrals in relative coordinates can be avoided, which is the case
in the $\delta$-function approximation. From our investigation we
note that this approximation is too drastic, as the valence nucleon
wave function has appreciable variation across the size of the $\alpha$
particle. 

\bibliographystyle{unsrt}

\end{document}